\documentclass[sigconf,nonacm]{acmart}

\usepackage{amsmath,booktabs}
\usepackage{graphicx}
\usepackage{tikz}
\usetikzlibrary{arrows.meta,positioning,calc}
\setcopyright{none}
\acmDOI{}
\acmISBN{}
\authorsaddresses{}
\acmConference[ICAIF '26]
  {ACM International Conference on AI in Finance}
  {November 14--17, 2026}
  {Milan, Italy}

\acmYear{2026}
\copyrightyear{2026}

\newcommand{\dd}{\mathrm{d}}
\newcommand{\E}{\mathbb{E}}

\newcommand{\relatedheading}[1]{%
  \par\noindent\textbf{#1}\par\nobreak\noindent\ignorespaces
}

\makeatletter

\def\@mkauthors@iii{%
  \global\setbox\mktitle@bx=\vbox{%

    \noindent\unvbox\mktitle@bx\par

    \vspace{8pt}

    \noindent
    \begin{minipage}[t]{0.47\textwidth}
      \centering
      \vspace{0pt}

      {\normalfont
       \fontsize{12.5pt}{14.5pt}\selectfont
       Siu Tung Wong\textsuperscript{1}\par}

      \vspace{3pt}

      {\normalfont
       \fontsize{10.5pt}{12pt}\selectfont
       \textsuperscript{1}Institute of Finance and Technology\par
       University College London\par
       London, United Kingdom\par}

      \vspace{2pt}

      {\normalfont
       \fontsize{10.5pt}{12pt}\selectfont
       \href{mailto:ucesstw@ucl.ac.uk}{ucesstw@ucl.ac.uk}\par}

    \end{minipage}
    \hfill
    \begin{minipage}[t]{0.47\textwidth}
      \centering
      \vspace{0pt}

      {\normalfont
       \fontsize{12.5pt}{14.5pt}\selectfont
       Carlo Campajola\textsuperscript{1,2}\par}

      \vspace{3pt}

      {\normalfont
       \fontsize{10.5pt}{12pt}\selectfont
       \textsuperscript{1}Institute of Finance and Technology\par
       University College London\par
       London, United Kingdom\par

       \vspace{2pt}

       \textsuperscript{2}UZH Blockchain Center\par
       Z\"urich, Switzerland\par}

      \vspace{2pt}

      {\normalfont
       \fontsize{10.5pt}{12pt}\selectfont
       \href{mailto:c.campajola@ucl.ac.uk}
            {c.campajola@ucl.ac.uk}\par}

    \end{minipage}

    \vspace{6pt}

    \centering
    {\normalfont
     \fontsize{10.5pt}{12pt}\selectfont
     Accepted for publication at ICAIF 2026.\par}

    \vspace{10pt}
  }%
}

\makeatother

\begin{document}

\title{When a Correct Reward Is Not Enough: Diagnosing and Guiding PPO in an
Analytically Solved Broker--Trader Game}

% ------------------------------------------------------------
% Author metadata
%
% These records are retained so that acmart / PDF metadata
% correctly identifies the authors and affiliations.
% ------------------------------------------------------------

\author{Siu Tung Wong}

\affiliation{%
  \department{Institute of Finance and Technology}
  \institution{University College London}
  \streetaddress{Gower Street}
  \city{London}
  \postcode{WC1E 6BT}
  \country{United Kingdom}
}

\email{ucesstw@ucl.ac.uk}

\author{Carlo Campajola}

\affiliation{%
  \department{Institute of Finance and Technology}
  \institution{University College London}
  \streetaddress{Gower Street}
  \city{London}
  \postcode{WC1E 6BT}
  \country{United Kingdom}
}

\affiliation{%
  \institution{UZH Blockchain Center}
  \streetaddress{Andreasstrasse 15}
  \city{Z\"urich}
  \postcode{8050}
  \country{Switzerland}
}

\email{c.campajola@ucl.ac.uk}

\renewcommand{\shortauthors}{Wong and Campajola}

\begin{abstract}
Reinforcement learning (RL) is increasingly used for financial optimal-control
problems when complex dynamics make analytical strategies difficult to obtain.
There are financial mathematics literactures which provides many solved models whose equations and
controls could evaluate and guide learning; we ask whether RL can exploit these
results.

We place a proximal policy optimisation (PPO) agent in an analytically solved
continuous-time broker--trader game. PPO replaces the broker and chooses its
trading speed while interacting with an informed trader and stochastic
uninformed order flow. 
We derive a finite-step reward from the broker's continuous-time payoff and
verify its discrete implementation through grid refinement and an exact
one-step identity. With zero uninformed flow, a validation-selected PPO--FFNN
approaches the reference action. With stochastic uninformed flow, the tested
PPO--FFNN and PPO--LSTM remain inaccurate, although supervised learning
confirms that their actors can represent the action. Monte Carlo diagnostics
show that their critics do not reliably rank nearby actions; potential-based
reward shaping also gives no reliable improvement. Under partial information,
a causal certainty-equivalent controller based on the broker's observable
history remains close to the reference, while PPO has larger errors and lower
payoffs.

Finally, we freeze the analytical policy and train PPO to adjust it after the
execution cost changes. Halving the cost yields a repeatable improvement that
closes \(2.22\%\) of the gap to the changed-cost reference. The analytical
solution therefore provides both a benchmark for diagnosing RL and a useful
starting policy for adaptation.
\end{abstract}



\maketitle

\section{Introduction}
\label{sec:introduction}

Optimal execution balances execution costs against inventory exposure and
adverse price movements. Reinforcement learning (RL) is attractive when richer
market dynamics make analytical strategies difficult to obtain. We ask whether
RL can recover an analytically known broker strategy from its performance
criterion and use that strategy after the environment changes.

Our environment follows the continuous-time game of Cartea, Jaimungal, and
S\'anchez-Betancourt \cite{cartea2025nashequilibriumbrokerstraders}, who model
the interaction between a broker, an informed trader, and an uninformed trader.
The broker internalises client orders and hedges in the lit market. Informed
flow contains private information, while rapid trading is costly and slow
trading leaves inventory exposed. Their analytical equilibrium gives an action
at every state, allowing same-state action comparisons together with realised
payoffs and inventory trajectories.

Prior studies use RL to adjust an Almgren--Chriss schedule, learn liquidation
policies against analytical benchmarks, or structure actor--critic learning
with closed-form policy and value functions
\cite{hendricks2014reinforcement,macri2024reinforcement,
wang2026reinforcement}. They do not test whether a finite-step reward recovers
the known state-dependent strategy in this broker--trader game. Analytical-policy
initialisation and adaptation have also been proposed
\cite{ning2021double,wang2026reinforcement}. We separate discretisation,
policy learning, and hidden-state estimation.

\relatedheading{We study three research questions:}
\begin{itemize}
\setlength{\topsep}{2pt}
\setlength{\partopsep}{0pt}
\setlength{\itemsep}{0pt}
\setlength{\parsep}{0pt}

\item \textbf{RQ1: Full-information control.}
Can PPO--FFNN and PPO--LSTM recover the broker's analytical trading strategy
when every relevant state variable is observed?

\item \textbf{RQ2: Partial-information control.}
How closely can PPO--FFNN and PPO--LSTM approach the certainty-equivalent
benchmark \cite{mania2019certainty} when the informed trader's inventory,
signal, and current trading rate are hidden?

\item \textbf{RQ3: Adapting an analytical policy.}
When the execution cost differs from the value used to calculate the analytical
policy, can PPO learn an adjustment that improves the unchanged policy?

\end{itemize}

Section~2 reviews related work, Section~3 introduces the environment,
Section~4 answers RQ1--RQ2, and Section~5 answers RQ3.
\section{Related Work}
\label{sec:related-work}

\relatedheading{Analytical execution and broker games:}
Analytical execution models relate trading actions to inventory, impact, order
flow, and predictive signals \cite{cartea2016orderflow,casgrain2019trading}.
Broker--trader games add strategic interaction because hedging affects prices
while client orders change the broker's inventory and information
\cite{cartea2025nashequilibriumbrokerstraders,wu2024brokertrader}. We use the
known broker action as the RL target.

\relatedheading{RL for optimal execution:}
RL has learned liquidation schedules through value-based methods, PPO, and
continuous-time actor--critic
\cite{nevmyvaka2006reinforcement,lin2020endtoend,ning2021double,
moallemi2022reinforcement,koulouris2026memory}. Hybrid methods adjust or
initialise from an analytical schedule, or embed closed-form structure
\cite{hendricks2014reinforcement,ning2021double,
wang2026reinforcement}. The broker--trader equilibrium instead permits
same-state action comparisons. Although policy gradients have convergence
results for linear--quadratic control \cite{fazel2018global}, PPO remains
sensitive to implementation and evaluation choices
\cite{engstrom2020implementation,andrychowicz2020what,henderson2018deep}.

\relatedheading{Analytical guidance and reward shaping:}
Analytical equations can guide learning and provide a benchmark. Prior work
adds a Hamilton--Jacobi--Bellman loss or closed-form policy and value structure
\cite{mukherjee2023bridging,wang2026reinforcement}. Residual RL adds a learned
action to an existing controller \cite{johannink2019residual}; in Section~5,
we use it to adjust the frozen broker action after the execution cost changes.
We also test potential-based shaping, which preserves policy ordering
\cite{ng1999policy,sowerby2022designing}.

\relatedheading{RL under partial information:}
With hidden state, an agent can summarise observations recurrently or estimate
the hidden variables explicitly \cite{igl2018deep,lambrechts2022recurrent}.
We compare feedforward and recurrent PPO with a certainty-equivalent controller
that updates such an estimate from observable history.
\section{Environment Design}
\subsection{Broker Execution Game}
\label{sec:broker-game}

Our baseline is the continuous-time broker--trader model of
Cartea, Jaimungal, and S\'anchez-Betancourt
\cite{cartea2025nashequilibriumbrokerstraders}. An informed trader sends orders
to the broker at rate \(\eta_t\), uninformed clients generate flow \(\xi_t\),
and the broker hedges the resulting inventory in a lit market at rate
\(\nu_t\). The informed trader responds to a predictive signal process
\(\alpha=(\alpha_t)_{t\in[0,T]}\). PPO chooses the broker's trading rate
\(\nu_t\), while the informed trader follows its analytical feedback rule and
the uninformed flow remains exogenous.

Let \(Q_t^B\) and \(Q_t^I\) denote broker and informed-trader inventories,
\(Y_t\) the price-impact state, and \(S_t\) the midprice. Their dynamics are
\begin{align}
 \dd Q_t^B&=(\nu_t-\eta_t-\xi_t)\dd t,
 &\dd Q_t^I&=\eta_t\dd t, \notag\\
 \dd Y_t&=(h\nu_t-pY_t)\dd t,
 &\dd\alpha_t&=-\kappa_\alpha\alpha_t\dd t
                  +\sigma_\alpha\dd W_t^\alpha, \notag\\
 \dd\xi_t&=-\kappa_\xi\xi_t\dd t+\sigma_\xi\dd W_t^\xi,
 &\dd S_t&=(\alpha_t+h\nu_t-pY_t)\dd t+\sigma_S\dd W_t^S .
 \label{eq:model-dynamics}
\end{align}
The broker's trade changes inventory, incurs execution cost, and affects the
price at which its remaining inventory is valued.

The analytical solution gives the feedback controls
\begin{align}
 \nu_t^\star
 &=P_{11}(t)Q^B_t+P_{12}(t)Q^I_t+P_{13}(t)Y_t
   +g_1(t)\alpha_t+g_2(t)\xi_t, \label{eq:nu-star}\\
 \eta_t^\star
 &=P_{21}(t)Q^B_t+P_{22}(t)Q^I_t+P_{23}(t)Y_t
   +h_1(t)\alpha_t+h_2(t)\xi_t. \label{eq:eta-star}
\end{align}
The time-dependent coefficients $P_{ij}(t)$, $g_i(t)$ and $h_i(t)$ follow from the Riccati equations in
\cite{cartea2025nashequilibriumbrokerstraders}. In the simulator, the informed
trader applies Equation~\eqref{eq:eta-star} at every state generated by the
broker's actions. Therefore even if the PPO agent takes suboptimal-actions,
the informed trader continues to use the analytical feedback
rule and does not operate in subgame nash equilibrium. We refer to it as the analytical-feedback trader.

\begin{figure*}[t]
\centering
\resizebox{0.98\textwidth}{!}{%
\begin{tikzpicture}[
  >=Latex,
  panel/.style={draw=black!38, rounded corners=3pt, fill=black!1,
    minimum width=54mm, minimum height=55mm},
  item/.style={draw=blue!55!black, rounded corners=2pt, fill=blue!5,
    text width=45mm, minimum height=9mm, align=center, inner sep=2mm},
  hidden/.style={item, draw=red!65!black, fill=red!6},
  network/.style={item, draw=violet!70!black, fill=violet!7,
    text width=44mm},
  reference/.style={item, dashed, draw=black!55, fill=black!3,
    text width=19mm, minimum height=6mm, inner sep=0.5mm},
  market/.style={item, draw=green!45!black, fill=green!7,
    text width=45mm},
  heading/.style={font=\bfseries\small, align=center},
  observed/.style={->, line width=0.85pt, draw=blue!60!black},
  fullonly/.style={->, dashed, line width=0.85pt, draw=red!65!black},
  process/.style={->, line width=1pt, draw=black!75},
  feedback/.style={->, rounded corners=4pt, dashed, line width=0.9pt,
    draw=green!45!black}
]
\node[panel] (information) {};
\node[panel, right=8mm of information] (policy) {};
\node[panel, right=8mm of policy] (execution) {};

\node[heading, anchor=north] at ([yshift=-2mm]information.north)
  {1. Information before decision \(k\)};
\node[item, anchor=north] (own) at ([yshift=-10mm]information.north)
  {\textbf{Broker and market state}\\
   \(S_k,Q_k^B,Y_k,\xi_k,\tau_k\) are observed;\\
   \(\nu_{k-1}\) also enters the partial policy};
\node[hidden, below=2mm of own] (informed)
  {\textbf{Informed trader and signal}\\
   \(Q_k^I,\alpha_k\) determine \(\eta_k^\star\); full-state only};
\node[item, below=2mm of informed] (uninformed)
  {\textbf{Uninformed clients}\\current flow \(\xi_k\) is observed};

\node[heading, anchor=north] at ([yshift=-2mm]policy.north)
  {2. Broker policy};
\node[item, anchor=north, text width=47mm] (observation)
  at ([yshift=-10mm]policy.north)
  {\textbf{Observation vector}\\[-1mm]
   full: \(o_k^F=(S,Q^B,Q^I,Y,\alpha,\xi,\tau)_k\)\\
   partial: \(o_k^P=(S,Q^B,Y,\xi,\tau,\nu_{k-1})_k\)};
\node[network, below=2mm of observation] (actor)
  {\textbf{Broker policy}\\PPO--FFNN / PPO--LSTM /\\
   analytical policy + PPO adjustment};
\node[reference, below=1.2mm of actor, xshift=-11.5mm] (referenceaction)
  {\textbf{Reference } \(\nu_k^{\mathrm{ref}}\)\\[-0.5mm]
   {\scriptsize evaluation only}};
\node[network, below=1.2mm of actor, xshift=11.5mm, text width=19mm,
  minimum height=6mm, inner sep=0.5mm] (action)
  {\textbf{Executed } \(\nu_k\)\\[-0.5mm]
   {\scriptsize clipped actor action}};

\node[heading, anchor=north] at ([yshift=-2mm]execution.north)
  {3. Execution and feedback};
\node[market, anchor=north] (litmarket) at ([yshift=-10mm]execution.north)
  {\textbf{Lit-market execution}\\
   \(\nu_k\) creates price impact \(h\nu_k\);
   client flows update \(Q^B\)};
\node[market, below=2mm of litmarket] (transition)
  {\textbf{Environment transition}\\
   produces \(S_{k+1},Q_{k+1}^B,Y_{k+1}\) and reward \(r_k\)};
\node[market, below=2mm of transition] (nextresponse)
  {updated state determines the next informed response
   \(\eta_{k+1}^{\star}\)};

\coordinate (obsmerge) at
  ($(own.east)!0.55!(observation.west |- own.east)$);
\draw[line width=0.85pt, draw=blue!60!black]
  (own.east) -- (obsmerge);
\draw[line width=0.85pt, draw=red!65!black, dashed]
  (informed.east) -|
  node[pos=0.25, above, font=\scriptsize]{full state} (obsmerge);
\draw[line width=0.85pt, draw=blue!60!black]
  (uninformed.east) -| (obsmerge);
\draw[observed] (obsmerge) -- (observation.west |- own.east);
\draw[process] (observation) -- (actor);
\draw[process] (actor) -- (action);
\draw[process] (action.east) -- (litmarket.west);
\draw[process] (litmarket) -- (transition);
\draw[process] (transition) -- (nextresponse);
\coordinate (feedbackright) at ($(nextresponse.south)+(0,-4mm)$);
\coordinate (feedbackleft) at ($(information.south west)+(-4mm,-3.5mm)$);
\draw[feedback] (nextresponse.south) -- (feedbackright) --
  node[midway, below, font=\scriptsize]
  {next decision; past \(\eta_k\) is now reconstructible}
  (feedbackleft) |- (informed.west);
\end{tikzpicture}}
\caption{Information and feedback for one broker decision. The policy maps the
full- or partial-information observation to a trading rate, which is clipped to
\([-100,100]\) before the environment updates the state and reward. Red
variables are hidden under partial information; past informed flow \(\eta_k\)
is reconstructible at decision \(k+1\). The reference action (\(\nu_k^{\mathrm{ref}}\)) is evaluated at every
PPO state.}
\Description{A three-column diagram shows the broker's observed and hidden
information, the observation entering an actor, the actor action driving the
environment, and the reference action \(\nu_k^{\mathrm{ref}}\) recalculated at
the state reached by PPO.
}
\label{fig:broker-information-flow}
\end{figure*}
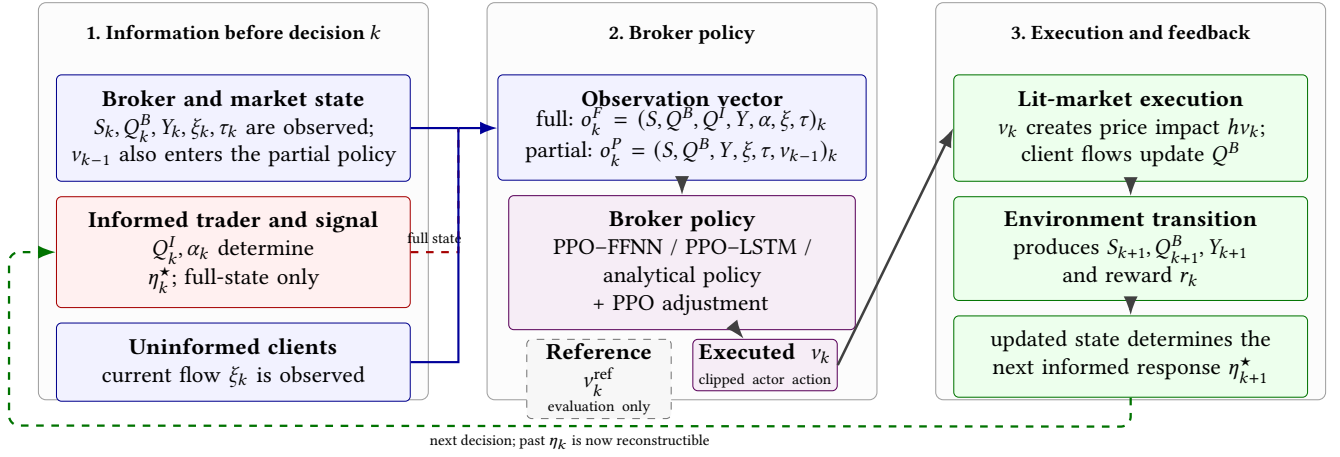

The environment uses Gymnasium 1.3.0 and PPO--FFNN is
implemented with Stable-Baselines3 2.9.0 and PPO--LSTM with SB3-Contrib 2.9.0.
The main parameters are \(T=1\), \(\Delta t=2\times10^{-4}\),
\(a=0.0012\), \(b=c=h=0.001\), \(p=r_B=r_I=0\),
\(\varphi=\psi=1\), \(\kappa_\alpha=5\), \(\sigma_\alpha=\sigma_S=1\),
and \(\kappa_\xi=15\), with \(\sigma_\xi\in\{0,100\}\). Initially
\(S_0=100\) and all other states are zero. Here \(a\) is the broker's
lit-market execution-cost coefficient,
while \(b\) and \(c\) determine the transaction prices for informed and
uninformed client flow; \(h,p\) are impact and resilience; \(r_B,r_I\)
are running inventory penalties; and \(\varphi,\psi\) are terminal inventory
penalties. Thus \(p=0\) gives zero resilience and \(r_B=0\) removes the
broker's running, but not terminal, inventory penalty.

\begin{table*}[t]
\caption{Networks and training settings. The reported zero-flow FFNN uses an
initial log standard deviation of 3; the other FFNN and LSTM experiments use
0. Checkpoints are selected by mean payoff on 50 separate validation paths.}
\label{tab:learning-configurations}
\centering
\scriptsize
\renewcommand{\arraystretch}{1.03}
\setlength{\tabcolsep}{3.2pt}
\begin{tabular}{@{}p{0.12\textwidth}p{0.25\textwidth}p{0.25\textwidth}
                    p{0.31\textwidth}@{}}
\toprule
Setting & PPO--FFNN & PPO--LSTM & Analytical policy with PPO adjustment \\
\midrule
Actor / critic
& Actor and critic each use two fully connected 64-unit ReLU hidden layers
& Actor and critic each use 24-unit LSTM and a fully
  connected 64-unit Tanh hidden layer
& Fixed analytical action; two-layer, width-64 Tanh correction network;
  two-layer, width-128 Tanh critic \\
PPO
Section~4 use
& Block~B: \(\xi=0\), log std \(3\); \(\sigma_\xi=100\), log std \(0\).
  Block~C: supervised, \(\xi=0\).
  Block~D: \(\sigma_\xi=100\), log std \(0\)
& Blocks~B and D: log std \(0\).
  Block~C: supervised, \(\xi=0\)
& Not used in Section~4 \\
& \multicolumn{2}{p{0.50\textwidth}}{
  Adam; rate \(10^{-4}\); \(\gamma=1\); GAE \(\lambda=0.99\);
  clip \(0.2\); batch 250; ten epochs; value/entropy coefficients
  \(0.5/0\); gradient norm \(0.5\); normalised advantages; state-dependent
  exploration sampled every four steps}
& 100 critic-only warm-up and 800 joint updates; four environments
  \(\times\) 1,024 steps; actor/critic rates \(2\!\times\!10^{-5}/
  2\!\times\!10^{-4}\); \(\gamma=1\), advantage-estimation
  \(\lambda=0.95\), clip \(0.1\), batch 1,024, ten epochs \\
Training / selection
& \multicolumn{2}{p{0.50\textwidth}}{
  \(4\times10^6\) transitions; five training seeds; activation and
  checkpoints selected by finite-grid payoff on separate validation paths}
& Five training seeds; checkpoints at updates
  50, 100, 200, 400, and 800; the highest mean payoff on 50 common
  validation paths is selected;
  final evaluation on 100 previously unused paths \\
Supervised diagnostic
& \multicolumn{2}{p{0.50\textwidth}}{
  ReLU FFNN and Tanh LSTM with the same actor shapes as PPO;
  Adam at \(10^{-3}\), 100 epochs; 40/20/40 analytical
  train/validation/test episodes}
& Not used \\
\bottomrule
\end{tabular}
\Description{A compact comparison of the actor and critic architectures,
PPO settings, training procedures, and analytical prior used by
the feedforward, recurrent, and analytical-policy-adjustment controllers.}
\end{table*}

Let \(X_t^B\) be the broker's cash. The broker maximises
\begin{equation}
 J^B(\nu)=\E\!\left[X_T^B+Q_T^BS_T-\varphi(Q_T^B)^2
 -r_B\int_0^T(Q_t^B)^2\dd t\right].
 \label{eq:broker-objective}
\end{equation}
After substituting the cash and price dynamics, this equals an
action-independent constant plus \(\E\int_0^T g_t^B\dd t\), with
\begin{align}
 g_t^B={}&-a\nu_t^2+b\eta_t^2+c\xi_t^2 \notag\\
 &+Q_t^B\!\left[
 \alpha_t+h\nu_t-pY_t
 -2\varphi(\nu_t-\eta_t-\xi_t)-r_BQ_t^B\right].
 \label{eq:broker-integrand}
\end{align}
Thus \(J^B\) is the broker's total expected payoff, while \(g_t^B\) is its
action-dependent payoff rate during trading.
 \subsection{Discrete PPO Environment}
\label{sec:rl-environment}

We divide \([0,T]\) into \(N=T/\Delta t\) intervals and use left-endpoint
Euler--Maruyama transitions \cite{higham2001algorithmic}.  Define
\(u_k=\nu_k-\eta_k-\xi_k\),
\(\mu_k=\alpha_k+h\nu_k-pY_k\),
\(\rho_\alpha=1-\kappa_\alpha\Delta t\), and
\(\rho_\xi=1-\kappa_\xi\Delta t\). The finite-grid dynamics are
\begin{align}
 Q_{k+1}^B&=Q_k^B+u_k\Delta t,
 &Q_{k+1}^I&=Q_k^I+\eta_k\Delta t, \notag\\
 Y_{k+1}&=Y_k+(h\nu_k-pY_k)\Delta t,
 & \alpha_{k+1}&=\rho_\alpha\alpha_k+\sigma_\alpha\Delta W_k^\alpha\\
 \xi_{k+1}&=\rho_\xi\xi_k+\sigma_\xi\Delta W_k^\xi ,
 &S_{k+1}&=S_k+\mu_k\Delta t+\sigma_S\Delta W_k^S, \notag.
 \label{eq:discrete-model-dynamics}
\end{align}
The three \(\Delta W\) terms are independent Gaussian increments with variance
\(\Delta t\). We define \(\tau_k=(T-t_k)/T\) as the fraction of the trading
horizon remaining. Full-information PPO receives
\(S_k\), \(Q_k^B\), \(Q_k^I\), \(Y_k\), \(\alpha_k\), \(\xi_k\), and
\(\tau_k\). Partial-information PPO receives \(S_k\), \(Q_k^B\), \(Y_k\),
\(\xi_k\), \(\tau_k\), and \(\nu_{k-1}\), while \(Q_k^I\) and \(\alpha_k\)
are hidden. Before entering either network, \(S\) is centred at 100, \(Y\) is
multiplied by 1,000, and \(\xi\) and the previous action are divided by 100; 
we found these rescaling provides better results as they place the network inputs on comparable numerical scales;
the dynamics and reported payoffs remain in the model's original units.

After the action has been executed, the broker observes \(Q_{k+1}^B\) and can
reconstruct the informed flow from the inventory transition:
\begin{equation}
 \eta_k=\nu_k-\xi_k-\frac{Q_{k+1}^B-Q_k^B}{\Delta t}.
 \label{eq:delayed-flow-reconstruction}
\end{equation}
The broker therefore cannot use \(\eta_k\) when choosing \(\nu_k\), but can use
the reconstructed value when forming its estimates for decision \(k+1\).

\subsection{Finite-step reward}

Let \(\Delta M_k^S=\sigma_S\Delta W_k^S\).
The transitions used in the reward calculation are
\begin{align}
 \Delta Q_k^B&=u_k\Delta t,
 \;\;\;\;\;\Delta S_k=\mu_k\Delta t+\Delta M_k^S, \notag\\
 \Delta X_k^B
 &=\left(-S_ku_k-a\nu_k^2+b\eta_k^2+c\xi_k^2\right)\Delta t .
 \label{eq:reward-transitions}
\end{align}
Applying the exact discrete product rule gives
\begin{align}
 \Delta(Q^BS)_k
 &=Q_k^B\Delta S_k+S_k\Delta Q_k^B
   +\Delta Q_k^B\Delta S_k \notag\\
 &=(Q_k^B\mu_k+S_ku_k)\Delta t
   +u_k\mu_k\Delta t^2+Q_{k+1}^B\Delta M_k^S,
 \label{eq:discrete-product}
\end{align}
while the terminal-inventory term satisfies
\begin{equation}
 (Q_{k+1}^B)^2-(Q_k^B)^2
 =2Q_k^Bu_k\Delta t+u_k^2\Delta t^2 .
 \label{eq:discrete-inventory-square}
\end{equation}
Adding the cash increment cancels
\(-S_ku_k\Delta t+S_ku_k\Delta t\). 
Therefore, define the accumulated finite-grid objective and its one-step
increment by
\begin{align}
 \mathcal V_k
 &=X_k^B+Q_k^BS_k-\varphi(Q_k^B)^2
   -\sum_{j<k}r_B(Q_j^B)^2\Delta t, \notag\\
 \mathcal V_{k+1}-\mathcal V_k
 &=g_k^B\Delta t+
   \left(u_k\mu_k-\varphi u_k^2\right)\Delta t^2
   +Q_{k+1}^B\Delta M_k^S .
 \label{eq:one-step-reward-identity}
\end{align}
The \(\Delta t^2\) terms are the cross-products generated by the discrete
product rule and squared-inventory increment. The remaining term,
\(Q_{k+1}^B\Delta M_k^S\), is the change in inventory value caused by the
unpredictable price innovation. Its conditional expectation is zero, so the
reward supplied to PPO is
\begin{equation}
 r_k=g_k^B\Delta t+
 \left(u_k\mu_k-\varphi u_k^2\right)\Delta t^2 .
 \label{eq:corrected-dense-reward}
\end{equation}
We write \(J_{\Delta t}^B(\pi)=\E_\pi[\sum_{k=0}^{N-1}r_k]\). Summing
Equation~\eqref{eq:one-step-reward-identity} shows that this expected return
equals the expected cash-and-inventory objective in
Equation~\eqref{eq:broker-objective}, up to the fixed initial value. On one
realised path, the two accumulated quantities differ by the omitted price
innovations \(\sum_kQ_{k+1}^B\Delta M_k^S\). The analytical payoff is
undiscounted, so we set \(\gamma_{\mathrm{RL}}=1\).

 % Generated by scripts/export_step9_results.py; do not edit by hand.
\providecommand{\StepNineGridDtCoarse}{10^{-3}}
\providecommand{\StepNineGridRegretCoarse}{3.462\times 10^{-7}}
\providecommand{\StepNineGridActionRMSECoarse}{0.016835}
\providecommand{\StepNineGridDtMedium}{5\times 10^{-4}}
\providecommand{\StepNineGridRegretMedium}{8.681\times 10^{-8}}
\providecommand{\StepNineGridActionRMSEMedium}{0.008468}
\providecommand{\StepNineGridDtFine}{2\times 10^{-4}}
\providecommand{\StepNineGridRegretFine}{1.383\times 10^{-8}}
\providecommand{\StepNineGridActionRMSEFine}{0.003389}
% Wider-exploration FFNN: outputs/icaif_v4/v32_exploration_full/v1.0.1/
% evaluation/compiled/aggregate_results.csv. Means and 95% Student-t margins
% are over five independent training seeds on 250 final test paths.
\providecommand{\ActFullZeroFFNNLoss}{0.045\mathbin{\pm}0.031}
\providecommand{\ActFullZeroFFNNRMSE}{5.494\mathbin{\pm}2.163}
\providecommand{\ActFullZeroLSTMLoss}{0.688\mathbin{\pm}0.687}
\providecommand{\ActFullZeroLSTMRMSE}{16.304\mathbin{\pm}2.921}
% Exact five-seed vectors are stored with the exported results.

\subsection{Environment Step}

We use PPO as our main algorithm \cite{schulman2017proximal} and compare FFNN and LSTM
actors, each trained for four million environment transitions.
At time \(t_k\), the
policy samples a raw action \(\nu_k^\pi\) (``Raw'' denotes the
network output before clipping), while the environment executes
\(\nu_k=\operatorname{clip}(\nu_k^\pi,-100,100)\), as shown in
Figure~\ref{fig:broker-information-flow}. PPO evaluates the likelihood of the
raw action during training. We record how often clipping occurs and the size of
the adjustment between the raw and executed actions. The informed trader
calculates \(\eta_k^\star\) from the state at time \(k\) before either action
changes the state.
 
\section{Experimental Results}
\label{sec:experimental-results}

We now answer the research questions introduced in Section~1. Block~A verifies
the reference action \(\nu_k^{\mathrm{ref}}\) used throughout the experiments. Blocks B–C answer RQ1 
by testing full-information PPO, actor capacity under supervised learning, the 
PPO critic’s ability to rank nearby actions, and potential-based reward shaping. Block~D answers RQ2 by
comparing partial-information PPO with a certainty-equivalent (CE) controller.
Section~\ref{sec:optimality-informed-ppo} answers RQ3.

Table~\ref{tab:learning-configurations} gives the training settings; code and
fixed experiment configurations will be released after peer review. Each PPO configuration uses five
training seeds and four million transitions per seed. Because PPO performance
can deteriorate late in training, we evaluate checkpoints every 0.4 million
transitions on 50 validation paths and select the checkpoint with the highest
average broker payoff. After checkpoint selection, every trained policy is evaluated 
on the same 250 test paths, which are not used for training or checkpoint selection. The reported 95\% Student-\(t\) intervals are calculated across five
independently trained policies, all evaluated on the same test paths.

\subsection{Block A: Continuous and finite-grid alignment}
\label{sec:block-a}

Block~A tests whether the comparison with PPO is affected by discretising the
continuous-time game. The simulator applies
Equation~\eqref{eq:nu-star} at every step of the Euler--Maruyama simulator.
Let \(x_k=(Q_k^B,Q_k^I,Y_k,\alpha_k,\xi_k)^{\mathsf T}\). The finite-grid
linear--quadratic (LQ) action is \(K_kx_k\), where \(K_k\) is the time-dependent feedback vector
obtained from the backward recursion below. We use the reference action as shown in
Figure~\ref{fig:broker-information-flow}, \(\nu_k^{\mathrm{ref}}=\operatorname{clip}(K_kx_k,-100,100).\)
This gives a feasible reference in
the PPO action range; the unconstrained LQ action is \(K_kx_k\).

To solve the finite-grid problem, write the Euler transition and reward as
\begin{align}
 x_{k+1}&=A_kx_k+B_k\nu_k+\varepsilon_{k+1}, \notag\\
 r_k(x_k,\nu_k)
 &=x_k^{\mathsf T}\mathsf Q_kx_k
   +2\nu_k s_k^{\mathsf T}x_k
   +R_k\nu_k^2,
 \label{eq:finite-grid-lq-form}
\end{align}
where \(x_k=(Q_k^B,Q_k^I,Y_k,\alpha_k,\xi_k)^{\mathsf T}\). If the next-step
value has quadratic coefficient \(F_{k+1}\), moving one step backwards gives
\begin{align}
 H_k&=\mathsf Q_k+A_k^{\mathsf T}F_{k+1}A_k,&
 \beta_k&=s_k+A_k^{\mathsf T}F_{k+1}B_k,\notag\\
 \chi_k&=R_k+B_k^{\mathsf T}F_{k+1}B_k,&
 K_k&=-\beta_k^{\mathsf T}/\chi_k,\notag\\
 F_k&=H_k-\beta_k\beta_k^{\mathsf T}/\chi_k.&&
 \label{eq:finite-grid-recursion}
\end{align}
The complete value is \(V_k^\star(x)=x^{\mathsf T}F_kx+c_k\), where
\(F_N=c_N=0\). For noise covariance \(\Sigma_k\),
\(c_k=c_{k+1}+\operatorname{tr}(F_{k+1}\Sigma_k)\). The constant does not
change the optimal action but is used in the reward-shaping experiment below.
Since \(\chi_k<0\), \(\nu_k^{\mathrm{LQ}}=K_kx_k\) is the unique unconstrained
maximiser. Completing the square gives
\(J_{\Delta t}^{B,\star}-J_{\Delta t}^{B}(\pi)=
\E_\pi\sum_k(-\chi_k)(\nu_k-\nu_k^{\mathrm{LQ}}(x_k))^2\).
We estimate it by
\begin{equation}
 L_{\mathrm{LQ}}
 =
 \frac{1}{M}\sum_{i=1}^{M}\sum_{k=0}^{N-1}
 (-\chi_k)
 \left[
  \nu_{i,k}^{\mathrm{exec}}
  -\nu_k^{\mathrm{LQ}}(x_{i,k})
 \right]^2,
 \label{eq:lq-loss}
\end{equation}
where \(M\) is the number of test paths, and \(\nu_{i,k}^{\mathrm{exec}}\) is the action taken by the agent. We refer to this quantity as the LQ
loss.

We therefore replay the continuous analytical feedback and the finite-grid LQ
feedback on 250 common market paths at three time resolutions.
Table~\ref{tab:blocks-a-d-summary}(a) shows that the action RMSE and LQ loss both
decrease as the grid is refined. At the grid used for training,
\(\Delta t=2\times10^{-4}\), they are
\(\StepNineGridActionRMSEFine\) and \(\StepNineGridRegretFine\). The linear
dynamics and Gaussian innovations make every unconstrained reference action
Gaussian. If its mean and standard deviation at decision \(k\) are \(m_k\) and
\(s_k\), the union bound gives
\begin{equation}
 \Pr\!\left(\max_k|\nu_k^{\mathrm{LQ}}|>100\right)
 \leq\sum_k\!\left[
 \Phi\!\left(\frac{-100-m_k}{s_k}\right)
 +1-\Phi\!\left(\frac{100-m_k}{s_k}\right)\right].
 \label{eq:action-bound-probability}
\end{equation}
For the zero-flow reference this bound is \(1.07\times10^{-5}\), and no
crossing occurs on 500 additional simulated paths.

\begin{table*}[!t]
% Section 4 values are traced to outputs/analysis/icaif_v20/existing_evidence/
% v1.0.0 and outputs/analysis/icaif_v19/matched_relu_supervised_ffnn/v1.0.0.
\caption{Summary of Blocks A--D. Action RMSE is the root mean squared difference
between the broker action and the stated reference action (\(\nu_k^{\mathrm{ref}}\)). (a) compares the
continuous analytical action with \(\nu_k^{\mathrm{ref}}\) at three time steps. (b)
compares full-information PPO with \(\nu_k^{\mathrm{ref}}\) when \(\xi=0\).
(c) evaluates the supervised FFNN and LSTM on analytical test states and then
when their actions determine subsequent states.
(d) reports payoff, payoff shortfall from the clipped reference, and
action RMSE when \(\sigma_\xi=100\). In (b) and (d), the reference is
recalculated at each state reached by the controller. Brackets are 95\% intervals across five
training seeds; the LSTM entry 11.85--339.70 is the five-seed range. Higher
payoff is better; lower LQ loss, payoff shortfall, and RMSE are better.}
\label{tab:blocks-a-d-summary}
\centering
\begin{minipage}[t]{0.49\textwidth}
\centering
{\normalsize\bfseries (a) Block A: Grid alignment\par}\smallskip
{\scriptsize
\setlength{\tabcolsep}{2.4pt}
\begin{tabular*}{\linewidth}{@{\extracolsep{\fill}}rrrr@{}}
\toprule
\(\Delta t\) & Steps \(N\) & LQ loss
 & \shortstack{Action\\RMSE} \\
\midrule
\(\StepNineGridDtCoarse\) & 1,000
  & \(\StepNineGridRegretCoarse\) & \(\StepNineGridActionRMSECoarse\) \\
\(\StepNineGridDtMedium\) & 2,000
  & \(\StepNineGridRegretMedium\) & \(\StepNineGridActionRMSEMedium\) \\
\(\StepNineGridDtFine\) & 5,000
  & \(\StepNineGridRegretFine\) & \(\StepNineGridActionRMSEFine\) \\
\bottomrule
\end{tabular*}
}
\end{minipage}\hfill
\begin{minipage}[t]{0.49\textwidth}
\centering
{\normalsize\bfseries (b) Block B: Full information, \(\xi=0\)\par}\smallskip
{\scriptsize
\setlength{\tabcolsep}{2.4pt}
\begin{tabular*}{\linewidth}{@{\extracolsep{\fill}}lrr@{}}
\toprule
Policy & \shortstack{Unconstrained\\LQ diagnostic} & Action RMSE \\
\midrule
FFNN--ReLU & \(\ActFullZeroFFNNLoss\) & \(\ActFullZeroFFNNRMSE\) \\
LSTM--Tanh & \(\ActFullZeroLSTMLoss\) & \(\ActFullZeroLSTMRMSE\) \\
\bottomrule
\end{tabular*}
}
\end{minipage}

\begin{minipage}[t]{0.49\textwidth}
\centering
{\normalsize\bfseries (c) Block C: Direct action supervision\par}\smallskip
{\scriptsize
\setlength{\tabcolsep}{2.4pt}
\begin{tabular*}{\linewidth}{@{\extracolsep{\fill}}lrr@{}}
\toprule
Actor & \shortstack{Supervised test\\RMSE}
 & \shortstack{Supervised closed-loop\\RMSE} \\
\midrule
FFNN--ReLU & 2.31 [1.99,2.63] & 2.06 [1.27,2.85] \\
LSTM--Tanh & 2.62 [2.44,2.81] & 11.85--339.70 \\
\bottomrule
\end{tabular*}
}
\end{minipage}\hfill
\begin{minipage}[t]{0.49\textwidth}
\centering
{\normalsize\bfseries (d) \(\sigma_\xi=100\): feasible comparisons\par}\smallskip
{\scriptsize
\setlength{\tabcolsep}{1.4pt}
\begin{tabular*}{\linewidth}{@{\extracolsep{\fill}}llrrr@{}}
\toprule
Info. & Controller & \(J_{\Delta t}^B\) & Shortfall & RMSE \\
\midrule
Full & Clipped reference & 0.1303 & 0 & 0 \\
Full & FFNN--ReLU & \(-7.060\) & 7.191 & 20.449 \\
Full & LSTM--Tanh & \(-39.194\) & 39.324 & 31.284 \\
Partial & CE estimate & 0.1297 & 0.00061 & 0.732 \\
Partial & FFNN--ReLU & \(-21.680\) & 21.810 & 24.819 \\
Partial & LSTM--Tanh & \(-24.108\) & 24.238 & 26.585 \\
\bottomrule
\end{tabular*}
}
\end{minipage}
\end{table*}

A separate numerical check verifies
Equation~\eqref{eq:one-step-reward-identity} to floating-point precision.
Together, the grid, action-bound, and reward checks show that these numerical
effects are much smaller than the PPO errors reported below.

\subsection{Block B: Full-information PPO recovery}
\label{sec:block-b}
We first set \(\xi = 0\), the zero-uninformed-flow model also considered in
\cite{cartea2025nashequilibriumbrokerstraders}. We then set
\(\sigma_\xi=100\) to introduce stochastic uninformed flow. Five independently
initialised policies per architecture observe the full state and train for four
million transitions. We compare each action by the policy with \(\nu_k^{\mathrm{ref}}\).

\begin{figure*}[t]
\centering
\includegraphics[width=0.99\textwidth]
{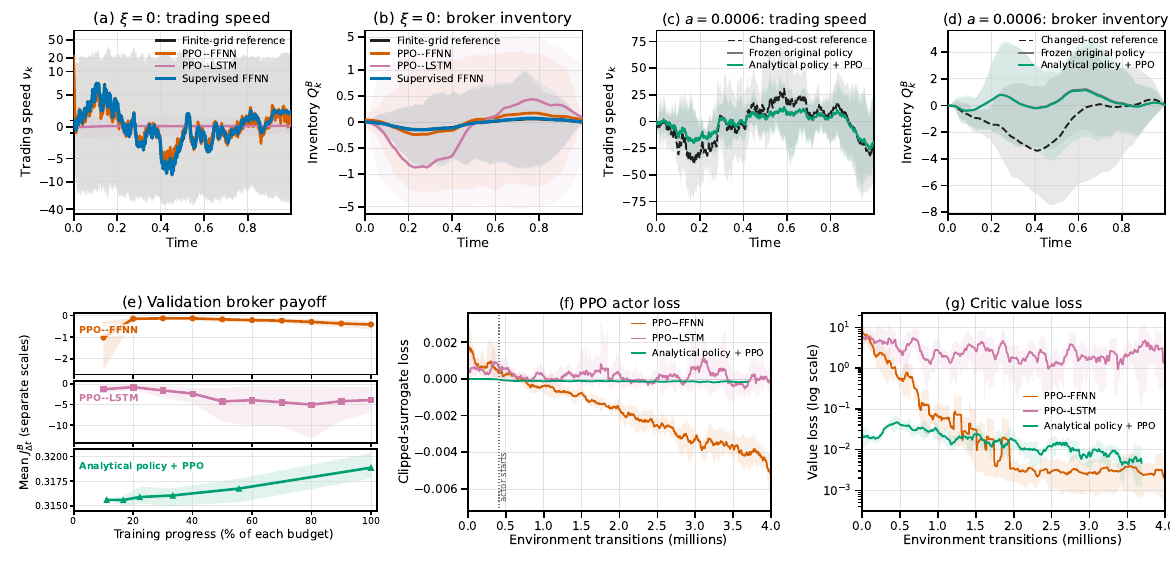}
\caption{Trajectories and PPO diagnostics. Panels (a)--(b) show the
zero-flow experiment and (c)--(d) the change from \(a=0.0012\) to \(0.0006\).
Lines are means and bands span the 5th--95th percentiles; (a)--(b) use signed
symmetric-log vertical scales. Panel (e) uses separate payoff scales and
normalises progress by each training budget; markers denote checkpoints.
Panels (f)--(g) show losses smoothed over 25 updates and averaged across five
training seeds. PPO--FFNN/LSTM are from Block~B and ``Analytical policy + PPO''
from Section~5. The dotted line marks the end of its 100-update critic warm-up.}
\Description{Seven panels show zero-flow broker trajectories, lower-cost
misspecification trajectories, validation broker payoff during learning, PPO
actor loss, and critic value loss.}
\label{fig:trajectory-comparisons}
\end{figure*}

For the zero-flow FFNN, we use an initial log standard deviation of \(3\).
Table~\ref{tab:blocks-a-d-summary}(b) reports an LQ loss of
\(\ActFullZeroFFNNLoss\) and an action RMSE of
\(\ActFullZeroFFNNRMSE\) across five independently trained roots.
Figure~\ref{fig:trajectory-comparisons} shows that the selected FFNN approaches
the zero-flow reference, whereas the LSTM remains inaccurate.
Table~\ref{tab:blocks-a-d-summary}(d) shows that the PPO agents tested with
\(\sigma_\xi=100\) also remain inaccurate. Thus, for RQ1, PPO--FFNN can
approach the full-information strategy when \(\xi=0\), but we do not obtain
similar performance after introducing stochastic uninformed flow.

\subsection{Block C: Actor capacity and learning diagnostics}
\label{sec:block-c}

Block~C tests whether the FFNN and LSTM can reproduce \(\nu_k^{\mathrm{ref}}\)
when that action is supplied directly. With \(\xi=0\), we train a ReLU FFNN
and a Tanh LSTM through supervised learning, using \(\nu_k^{\mathrm{ref}}\) as
labels and the supervised learning has the same architectures as the PPO actors. We evaluate them on
analytical test trajectories and then allow each actor to control the
simulator.

Table~\ref{tab:blocks-a-d-summary}(c) shows that supervised learning
substantially reduces the test error of both architectures. The supervised FFNN
also remains close to the reference when it controls the simulator, whereas the
LSTM result varies considerably across training seeds. Supervised learning therefore shows that both networks can approximate
\(\nu_k^{\mathrm{ref}}\) during testing. 
Figure~\ref{fig:trajectory-comparisons} shows the
performance of the models.

\relatedheading{Reward and critic diagnostics:}
Starting from the full-information baseline
(\(\sigma_\xi=100\), ReLU FFNN), we change one component at a time to probe
the sensitivity of PPO. The analytical broker action and corrected reward do
not depend on the midprice \(S_k\). We therefore remove \(S_k\) to test
whether this input makes PPO learning more difficult, and varify the effects of
the \(\Delta t^2\) correction by removing it or multiplying the reward by 100. Every
condition uses five seeds, four million training transitions per seed, and the
same numbers of validation and test paths. Table~\ref{tab:ppo-learning-diagnostics}
reports the resulting payoff, action RMSE, and critic explained variance.
Removing the \(\Delta t^2\) correction or the price input \(S_k\) does not
produce a consistent payoff change across seeds. Multiplying the reward by 100
lowers the payoff for every seed, showing that PPO is sensitive to reward scale.

\begin{table}[t]
% Values are traced to outputs/icaif_v4/v19_reward_observation/v1.0.0/
% evaluation/final_test/final_results.json.
\caption{Full-information ReLU FFNN ablations at \(\sigma_\xi=100\).
\(J_{\Delta t}^B\) is evaluated using Equation~\eqref{eq:corrected-dense-reward}, 
and action RMSE is calculated using \(\nu^{ref}_k\). Explained Variance (EV) is the 
critic's value predictions at the selected checkpoint. Relative to the corrected baseline, paired
payoff changes (95\% intervals) are \(1.48[-7.34,10.31]\) without the
\(\Delta t^2\) term, \(-29.24[-33.93,-24.54]\) after multiplying rewards by
100, and \(-1.09[-8.56,6.37]\) without \(S_k\). Higher payoff and EV, and lower
RMSE, are better.}
\label{tab:ppo-learning-diagnostics}
\centering
\footnotesize
\setlength{\tabcolsep}{2pt}
\begin{tabular*}{\columnwidth}{@{\extracolsep{\fill}}lrrr@{}}
\toprule
Training reward / input & \(J_{\Delta t}^B\) & Action RMSE & EV \\
\midrule
Corrected / with \(S_k\) & \(-5.75\!\pm\!4.82\) & \(19.76\!\pm\!4.15\) & \(.93\!\pm\!.07\) \\
First order / with \(S_k\) & \(-4.27\!\pm\!5.59\) & \(17.95\!\pm\!4.63\) & \(.79\!\pm\!.15\) \\
Corrected \(\times100\) / with \(S_k\) & \(-34.99\!\pm\!.78\) & \(31.01\!\pm\!.19\) & \(.42\!\pm\!.35\) \\
Corrected / without \(S_k\) & \(-6.84\!\pm\!7.79\) & \(19.13\!\pm\!5.66\) & \(.84\!\pm\!.15\) \\
\bottomrule
\end{tabular*}
\Description{Broker payoff and action error for PPO trained with the corrected
or first-order reward, with a fixed reward scaling, and without the absolute
price input, together with in-sample critic explained variance.}
\end{table}

PPO learns a state-value function \(V_\theta(s)\). Since
\(\gamma_{\mathrm{RL}}=1\), the Bellman relation gives
\[
 Q_k^\pi(s,a)
 =\mathbb E\!\left[r_k(s,a)+V_{k+1}^\pi(s_{k+1})
   \mid s_k=s,\nu_k=a\right].
\]
We test whether the critic correctly ranks two actions near the trained policy.
At a test state \(s\) that is not used to train or select the critic, let
\(\bar\nu_\pi(s)\) be the deterministic policy action and set
\(a_\pm=\operatorname{clip}(\bar\nu_\pi(s)\pm10,-100,100)\). Clipping is
inactive in these 120 comparisons, so the two actions are 20 units apart. For
Monte Carlo sample \(\ell\), \(r_k^\ell(s,a)\) is the immediate reward after
taking action \(a\), and \(s_{k+1}^{(a,\ell)}\) is the resulting next state.
Replacing \(V^\pi\) with \(V_\theta\) and averaging over \(L\) simulated
transitions gives
{\footnotesize
\begin{align}
 \Delta\widehat Q_{\theta,k}
 &=\frac1L\sum_{\ell=1}^{L}\Biggl[
 \underbrace{\Bigl(r_k^{\ell}(s,a_+)
 +V_\theta(s_{k+1}^{(a_+,\ell)})\Bigr)}_{
 \widehat Q_{\theta,k}^{\ell}(s,a_+)}
 -\underbrace{\Bigl(r_k^{\ell}(s,a_-)
 +V_\theta(s_{k+1}^{(a_-,\ell)})\Bigr)}_{
 \widehat Q_{\theta,k}^{\ell}(s,a_-)}
 \Biggr],\notag\\
 \Delta Q_k^{\mathrm{MC}}
 &=\frac1L\sum_{\ell=1}^{L}\bigl[
 \mathcal G_{k,\ell}^{\pi}(a_+)-
 \mathcal G_{k,\ell}^{\pi}(a_-)\bigr],
 \label{eq:critic-local-action-comparison}
\end{align}
}
where \(\mathcal G_{k,\ell}^{\pi}(a)
=
r_k^\ell(s,a)
+
\sum_{j=k+1}^{N-1}
r_j^\ell
\left(s_j^{a,\ell},\nu_j^\pi\right),\) is the sum of the remaining rewards from
Monte Carlo sampling after selecting
action \(a\) at decision
\(k\). 
\(\Delta\widehat Q_{\theta,k}\) is the action preference predicted
by the critic using the immediate reward and next-state value, while
\(\Delta Q_k^{\mathrm{MC}}\) is the action preference estimated from the
complete simulated returns (if \(\Delta\widehat Q_{\theta,k}\) and  \(\Delta Q_k^{\mathrm{MC}}\) have the same sign, it indicates that the
critic correctly ranks the two actions). The two candidate actions use the same
market innovations and future policy noise, and \(L=32\) Monte Carlo
simulations are averaged for each comparison.

\begin{figure*}[t]
\centering
\begin{tikzpicture}[baseline=(current bounding box.north)]
\node[inner sep=0, anchor=north west] (critic-full) at (0,0)
{\includegraphics[height=0.245\textwidth]
{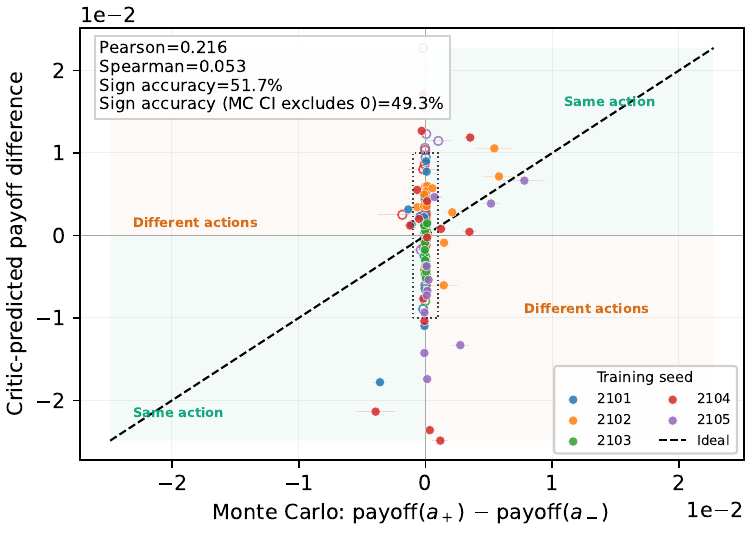}};
\node[inner sep=0, anchor=north west] (critic-zoom)
at ([xshift=1.5mm]critic-full.north east)
{\includegraphics[height=0.20\textwidth]
{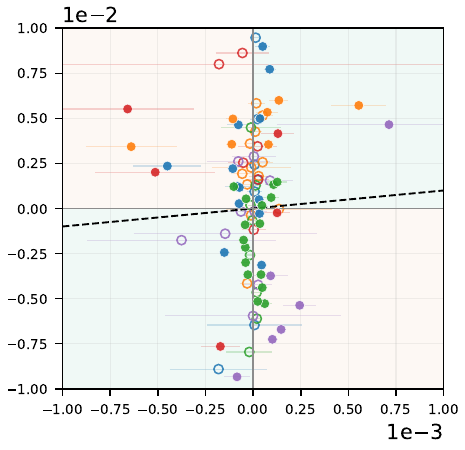}};
\node[inner sep=0, anchor=north west] (payoff-shortfall)
at ([xshift=1.5mm]critic-zoom.north east)
{\includegraphics[height=0.245\textwidth]
{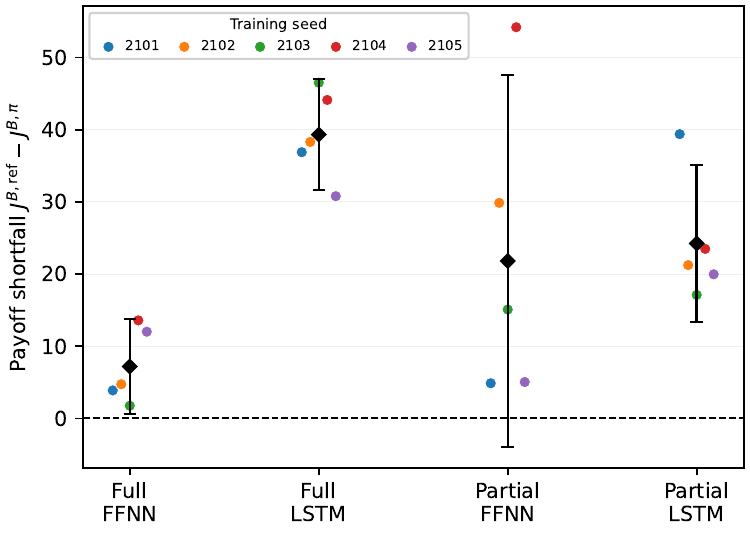}};
\draw[-{Latex[length=1.8mm]}, draw=black!55, line width=0.8pt]
([xshift=5mm,yshift=1mm]critic-full.center)
to[bend left=20]
node[pos=0.58, above, font=\scriptsize, text=black!65]{}
([xshift=-1mm,yshift=2mm]critic-zoom.west);
\end{tikzpicture}
\caption{Learning diagnostics. Left and centre: each point compares \(a_+\) and
\(a_-\) at one of 120 test states. The horizontal axis
is their difference in cumulative
reward estimated by Monte Carlo; the vertical axis is the difference predicted
using the immediate reward and critic value of the next state. Each point is
\((\Delta Q_k^{\mathrm{MC}},\Delta\widehat Q_{\theta,k})\), where both differences
subtract the payoff of \(a_-\) from \(a_+\). In the top-right region, both
estimates prefer \(a_+\); in the bottom-left region, both prefer \(a_-\). In the
top-left region, Monte Carlo prefers \(a_-\) but the critic predicts \(a_+\),
while in the bottom-right region, Monte Carlo prefers \(a_+\) but the critic
predicts \(a_-\). Pearson correlation measures whether the critic difference 
increases with the Monte Carlo difference, while Spearman correlation measures 
whether the critic ranks the action comparisons in the same order as Monte Carlo.
The pooled Pearson and Spearman correlations are 0.216 and 0.053, and sign
accuracy is 51.7\%.
Horizontal bars are 95\% Monte Carlo intervals; open points indicate decisions
where \(a_+\) and \(a_-\) appear equally good within this uncertainty. Without including open points
the remaining 71 points show a sign accuracy is 49.3\%. Right: each seed's payoff
shortfall from the clipped reference, with \(\sigma_{\xi} = 100\); diamonds and bars show the mean and 95\%
Student-\(t\) interval. Correlations closer to one and higher sign accuracy are
better; lower payoff shortfall is better.}
\Description{The left panel shows weak association between critic and Monte
Carlo local action preferences, the centre panel enlarges the dense region near
zero, and the right panel shows five seed-level payoff shortfalls for full- and
partial-information FFNN and LSTM policies.}
\label{fig:learning-diagnostics}
\end{figure*}

We also compare \(V_\theta(s)\) with the Monte Carlo return of the current
policy at 24 test states per seed. The critics can distinguish states with 
generally higher or lower expected remaining returns, but 
Figure~\ref{fig:learning-diagnostics} shows that they cannot reliably rank 
two nearby actions taken from the same state. Inaccurate local action ranking is
therefore a plausible source of weak PPO performance in \(\xi = 100\).

\relatedheading{Potential-based reward shaping:}
We next test whether the long-horizon broker payoff provides insufficient
information about the quality of each action. The potential
\(V_k^\star(x)=x^{\mathsf T}F_kx+c_k\) is computed from the finite-grid LQ
solution, which is used to shape our reward from Equation~\eqref{eq:corrected-dense-reward}. 
The PPO is trained with
\begin{align}
 r_k^{\mathrm{shape}}
 &=r_k+V_{k+1}^\star(x_{k+1})-V_k^\star(x_k),\notag\\
 \E[r_k^{\mathrm{shape}}\mid x_k,\nu_k]
 &=\chi_k\bigl(\nu_k-K_kx_k\bigr)^2\leq0.
 \label{eq:potential-shaped-reward}
\end{align}
Because \(V_N^\star=0\) and the initial state is fixed, the shaped and original
episode returns differ by the same constant. Reward shaping therefore changes
the learning signal without changing the ordering of policies
\cite{ng1999policy}. Since \(K_kx_k\) is calculated from the unconstrained finite-grid LQ problem
it can produce trading speeds that lie outside the action range,
the squared term measures deviation from the unconstrained LQ action.

We train ten FFNN policies, five with the original reward in
Equation~\eqref{eq:corrected-dense-reward} and 5 with the shaped reward in
Equation~\eqref{eq:potential-shaped-reward}. Each pair has the same initial
weights, market paths, and training settings. Shaping increases mean payoff by \(1.57\) (95\% interval [-5.00,8.14]) and reduces
action RMSE by \(1.76\) (95\% interval [-4.50,8.02]); both changes are
favourable in three of five seeds. It also increases the per-step reward
standard deviation by approximately 75\%. Raw training actions cross the
feasible range in approximately 0.1\% of decisions under either reward. Under
this protocol, shaping changes the learning signal but does not reliably
recover \(\nu_k^{\mathrm{ref}}\).

\subsection{Block D: Partial-information control}
\label{sec:block-d}
Block~D asks how closely PPO can approach the causal CE controller when both 
use the same information available before each decision. PPO observes
\((S,Q^B,Y,\xi,\tau,\nu_{k-1})\), while \(Q^I\), \(\alpha\), and the current
informed-trader rate are hidden before acting. The CE controller uses the game
dynamics and the broker's causal history
\[
 \mathcal H_k^B=\sigma(o_0^P,\nu_0,\ldots,\nu_{k-1},o_k^P)
\]
to estimate the hidden components \(Q_k^I\) and \(\alpha_k\). It then
substitutes these estimates into the finite-grid feedback rule:
\[
 \nu_k^{\mathrm{CE}}=\Pi_{\mathcal A}\!\left(K_k\widehat x_k\right).
\]
Specifically, the CE controller forms
\[
 \widehat Q^I_{k\mid k^-}
 =\mathbb E[Q_k^I\mid\mathcal H_k^B],
 \qquad
 \widehat\alpha_{k\mid k^-}
 =\mathbb E[\alpha_k\mid\mathcal H_k^B].
\]
After observing \(Q_{k+1}^B\), the controller reconstructs the informed flow
that has already occurred and updates the informed trader's inventory:
\begin{equation}
\begin{aligned}
 \eta_k^{\mathrm{rec}}
 &=\nu_k-\xi_k-\frac{Q_{k+1}^B-Q_k^B}{\Delta t},\\
 \widehat Q^I_{k+1\mid k+1^-}
 &=\widehat Q^I_{k\mid k^-}+\eta_k^{\mathrm{rec}}\Delta t .
\end{aligned}
\label{eq:ce-causal-update}
\end{equation}
Writing \(P_{ij,k}=P_{ij}(t_k)\) and \(h_{i,k}=h_i(t_k)\), we rearrange the
informed trader's feedback equation to estimate the signal:
\begin{equation}
\begin{aligned}
 \widehat\alpha_{k\mid k}
 &=\bigl(\eta_k^{\mathrm{rec}}-P_{21,k}Q_k^B
 -P_{22,k}\widehat Q^I_{k\mid k^-}\\
 &\hspace{8mm}-P_{23,k}Y_k-h_{2,k}\xi_k\bigr)/h_{1,k},\\
 \widehat\alpha_{k+1\mid(k+1)^-}
 &=\rho_\alpha\widehat\alpha_{k\mid k}.
\end{aligned}
 \label{eq:ce-signal-inversion}
\end{equation}
The estimated informed-trader inventory and predictive signal are combined 
with the observed state variables to calculate the CE action \(\nu_{k+1}^{\mathrm{CE}}\)
at the next decision. On 250 test paths, the
reconstructed informed-trader inventory has negligible error, while the
predicted signal and pre-action informed flow have RMSE 0.0141 and 0.873. The
remaining error is caused by the new signal innovation, which is unavailable
before the broker acts.

Table~\ref{tab:blocks-a-d-summary}(d) shows that the CE controller remains close
to \(\nu_k^{\mathrm{ref}}\), while both partial-information PPO agents have
substantially lower payoffs and larger action errors. The broker's observable
history therefore contains enough information for accurate control in this
environment, but the PPO agents do not learn to use that history as
effectively. This answers RQ2.
\section{Adapting the Analytical Policy with PPO}
\label{sec:optimality-informed-ppo}

We now address RQ3 by testing the direction we introduced in
Section~\ref{sec:introduction} and proposed in
\cite{ning2021double,wang2026reinforcement}: using an analytical policy as the
starting point for RL. Our propsed method would be to freeze the analytical broker action and train PPO to
correct execution-cost misspecification through a residual adjustment
\cite{johannink2019residual}.

Let \(\mathcal E(a)\) denote the broker environment with execution-cost
coefficient \(a\). We calculate the analytical policy in the original
environment
\(\mathcal E_0=\mathcal E(a_0)\), where \(a_0=0.0012\), and train the PPO
adjustment in
\(\mathcal E_{\mathrm{new}}=\mathcal E(a_{\mathrm{true}})\).

\relatedheading{Frozen analytical policy:}
In \(\mathcal E_0\), we evaluate and store the analytical coefficients
\(P_{11},P_{12},P_{13}\) and \(g_1,g_2\) from
\cite{cartea2025nashequilibriumbrokerstraders}, with the optimal action
\begin{equation}
 \nu_k^{\mathcal E_0}=
 P_{11}^0(t_k)Q_k^B+P_{12}^0(t_k)Q_k^I+P_{13}^0(t_k)Y_k
 +g_1^0(t_k)\alpha_k+g_2^0(t_k)\xi_k .
\label{eq:exact-analytical-prior}
\end{equation}
Equation~\eqref{eq:exact-analytical-prior} has the same broker-feedback form as
Equation~\eqref{eq:nu-star}, and the superscript \(0\) identifies coefficients
calculated in \(\mathcal E_0\). On 100 test paths generated at \(a_0\), the
% Recomputed on seeds 40043--40142; see outputs/analysis/icaif_v28/
% nominal_exact_policy_equivalence/v1.0.0/metrics.json in the experiment repo.
frozen and replayed analytical trajectories have trading-speed and inventory
RMSEs of \(6.36\times10^{-7}\) and \(3.29\times10^{-9}\), respectively.

\relatedheading{PPO adjustment after a cost change:}
In \(\mathcal E_{\mathrm{new}}=\mathcal E(a_{\mathrm{true}})\), the analytical
action calculated in \(\mathcal E_0\) remains fixed and PPO learns an additive
correction:
\begin{equation}
 \mu(o_k)
 =\nu_k^{\mathcal E_0}+\delta\nu(o_k) .
 \label{eq:analytical-residual-policy}
\end{equation}
Here, \(\nu_k^{\mathcal E_0}\) is the action generated by the frozen
analytical policy and \(\delta\nu(o_k)\) is the residual adjustment learned by PPO.
The residual is a two-layer, width-64 Tanh FFNN of the full-information
observation in \(\mathcal E_{\mathrm{new}}\) (we also tested the ReLU setup from Block~B, the wider exploration improves
learning when \(\xi=0\), but it lowers payoff for all five residual
policies; thus we found this Tanh setup to be the best for our misspecification problem). Its output layer is initialised at
zero, so the initial policy mean equals \(\nu_k^{\mathcal E_0}\). PPO then
trains the residual without an additional adjustment bound or penalty.
Gaussian exploration is centred on the combined mean, and the environment
clips sampled actions to \([-100,100]\). Table~\ref{tab:learning-configurations}
reports the remaining training settings.

All three cost experiments use full information and \(\sigma_\xi=100\). We
test \(a_{\mathrm{true}}\in\{0.0006,0.0012,0.0018\}\) in
\(\mathcal E_{\mathrm{new}}\). The informed trader's
coefficients are recomputed at the true cost, while the broker retains the
coefficients from \(\mathcal E_0\) and does not observe \(a_{\mathrm{true}}\) as an input. A
separate PPO adjustment is trained for each cost. For evaluation, \(\nu_k^{\mathrm{ref}}\) 
is recomputed at the true cost. For each seed, we evaluate five checkpoints on 50 validation paths and, as in
Section~4, select the checkpoint with the highest mean finite-grid broker
payoff. The selected policy is then evaluated on 100 common test paths.

\begin{table}[t]
% Values are traced to outputs/runs/unrestricted_analytical_residual_ppo_
% misspecification/unrestricted_section5_ablation_20260808/aggregate_by_cost.csv.
\caption{Finite-grid payoff after changing execution cost. ``Reference'' is
recalculated in \(\mathcal E_{\mathrm{new}}\), while ``frozen'' uses
\(\mathcal E_0\). Brackets are 95\% Student-\(t\) intervals over five training
seeds after averaging 100 common paths. Higher payoff and change are better.}
\label{tab:execution-cost-results}
\centering
\footnotesize
\setlength{\tabcolsep}{2pt}
\begin{tabular*}{\columnwidth}{@{\extracolsep{\fill}}rrrrr@{}}
\toprule
\(a_{\rm true}\) & Reference & Frozen & PPO adjusted & Change (95\% CI) \\
\midrule
0.0006 & 0.53010 & 0.29632 & 0.30151 & \(0.00519\;[0.00296,0.00743]\) \\
0.0012 & 0.06882 & 0.06882 & 0.06901 & \(0.00019\;[-0.00041,0.00079]\) \\
0.0018 & \(-0.15783\) & \(-0.16164\) & \(-0.16145\) & \(0.00019\;[-0.00034,0.00071]\) \\
\bottomrule
\end{tabular*}
\Description{Mean broker payoff for the changed-cost reference, unchanged
analytical policy, and PPO-adjusted policy at three execution costs.}
\end{table}

When the execution cost is halved, all five PPO adjustments improve the test
payoff. The mean increase is \(0.00519\), with a 95\% interval of
\([0.00296,0.00743]\) across the five independently trained policies, and
closes \(2.22\%\) of the gap between the unchanged policy and the changed-cost
reference. The largest residual action on the lower-cost test paths is 1.41,
and final-action clipping occurs on 0.029\% of decisions. PPO therefore provides a small,
repeatable improvement only in the lower-cost case tested here.
\section{Discussion and Conclusion}
\label{sec:conclusion}

We studied whether PPO can recover and subsequently adapt an analytically known
broker strategy. For RQ1, we have shown that PPO--FFNN approaches
\(\nu_k^{\mathrm{ref}}\) under full information when \(\xi=0\), however when 
\(\xi=100\), the policies remain inaccurate. The finite-step reward
and grid-refinement results show that these errors are not caused by numerical issues,
while supervised learning shows that the FFNN can
approximate and execute the reference action on states generated by the
analytical policy. The critic diagnostic shows inaccurate local
action ranking can contribute to the difficulty of learning from rewards.

For RQ2, the certainty-equivalent controller remains close to the reference 
action \(\nu_k^{\mathrm{ref}}\) when the informed trader's inventory and signal are
hidden. Its performance shows that the available
history supports accurate control in this environment, while PPO--FFNN and
PPO--LSTM do not learn to use that information as effectively.

For RQ3, the frozen analytical policy provides an accurate starting strategy
after the execution cost changes. Residual PPO produces a small and repeatable
payoff improvement when the cost is halved. The analytical
solution is therefore useful both as a benchmark for diagnosing learned
policies and as a prior for subsequent adaptation, thus providing
some evidence of analytical policy initialisation being useful \cite{ning2021double}.

The numerical checks and supervised learning make discretisation, reward
implementation, and insufficient FFNN capacity unlikely to be the primary
limitations. A more plausible explanation is the interaction between
exploration and reward-based value estimation. Wider initial exploration
improves PPO when \(\xi = 0\), but performance remains poor after
stochastic uninformed flow is introduced, where the critic also shows weak
local action ranking. Partial information additionally requires relevant
information to be recovered from past observations. When an analytical policy
is already available, wider exploration can instead move the adjustment away
from a useful starting point, while conservative adjustments produce a small
improvement under the lower-cost misspecification.

Future work can use analytical value or action-value information to improve the
local learning signal, study alternative action and policy parameterisations,
and develop structured residual policies for nonlinear costs or uncertain
market impact. Under partial information, it would be interesting to test the performance
of the model by introducing a filter into the actor, and practical applications for the partial
information setting would include having a \(\xi + \eta\) flow.
\bibliographystyle{ACM-Reference-Format}
\bibliography{references}

@article{cartea2025nashequilibriumbrokerstraders,
  author  = {Cartea, {\'A}lvaro and Jaimungal, Sebastian and S{\'a}nchez-Betancourt, Leandro},
  title   = {Nash Equilibrium between Brokers and Traders},
  journal = {Finance and Stochastics},
  year    = {2026},
  doi     = {10.1007/s00780-026-00595-7},
  url     = {https://doi.org/10.1007/s00780-026-00595-7},
  note    = {Published online 27 May 2026}
}

@article{cartea2016orderflow,
  author  = {Cartea, {\'A}lvaro and Jaimungal, Sebastian},
  title   = {Incorporating Order-Flow into Optimal Execution},
  journal = {Mathematics and Financial Economics},
  volume  = {10},
  number  = {3},
  pages   = {339--364},
  year    = {2016},
  doi     = {10.1007/s11579-016-0162-z},
  url     = {https://doi.org/10.1007/s11579-016-0162-z}
}

@article{casgrain2019trading,
  author  = {Casgrain, Philippe and Jaimungal, Sebastian},
  title   = {Trading Algorithms with Learning in Latent Alpha Models},
  journal = {Mathematical Finance},
  volume  = {29},
  number  = {3},
  pages   = {735--772},
  year    = {2019},
  doi     = {10.1111/mafi.12194},
  url     = {https://doi.org/10.1111/mafi.12194}
}

@misc{wu2024brokertrader,
  author        = {Wu, Xuchen and Jaimungal, Sebastian},
  title         = {Broker-Trader Partial Information Nash-Equilibria},
  year          = {2024},
  eprint        = {2412.17712},
  archivePrefix = {arXiv},
  primaryClass  = {q-fin.MF},
  doi           = {10.48550/arXiv.2412.17712},
  url           = {https://arxiv.org/abs/2412.17712}
}

@misc{koulouris2026memory,
  author        = {Koulouris, Christos Spyridon and Campajola, Carlo},
  title         = {Memory-Induced Supra-Competitive Outcomes Between Deep Reinforcement Learning Agents in Optimal Trade Execution},
  year          = {2026},
  eprint        = {2605.20348},
  archivePrefix = {arXiv},
  primaryClass  = {q-fin.CP},
  doi           = {10.48550/arXiv.2605.20348},
  url           = {https://arxiv.org/abs/2605.20348}
}

@article{ning2021double,
  author  = {Ning, Brian and Lin, Franco Ho Ting and Jaimungal, Sebastian},
  title   = {Double Deep Q-Learning for Optimal Execution},
  journal = {Applied Mathematical Finance},
  volume  = {28},
  number  = {4},
  pages   = {361--380},
  year    = {2021},
  doi     = {10.1080/1350486X.2022.2077783}
}

@article{lambrechts2022recurrent,
  author  = {Lambrechts, Gaspard and Bolland, Adrien and Ernst, Damien},
  title   = {Recurrent Networks, Hidden States and Beliefs in Partially Observable Environments},
  journal = {Transactions on Machine Learning Research},
  year    = {2022},
  url     = {https://openreview.net/forum?id=dkHfV3wB2l}
}

@inproceedings{igl2018deep,
  author    = {Igl, Maximilian and Zintgraf, Luisa M. and Le, Tuan Anh and Wood, Frank and Whiteson, Shimon},
  title     = {Deep Variational Reinforcement Learning for {POMDPs}},
  booktitle = {Proceedings of the 35th International Conference on Machine Learning},
  series    = {Proceedings of Machine Learning Research},
  volume    = {80},
  pages     = {2117--2126},
  year      = {2018},
  publisher = {PMLR},
  url       = {https://proceedings.mlr.press/v80/igl18a.html}
}

@misc{schulman2017proximal,
  author        = {Schulman, John and Wolski, Filip and Dhariwal, Prafulla and Radford, Alec and Klimov, Oleg},
  title         = {Proximal Policy Optimization Algorithms},
  year          = {2017},
  eprint        = {1707.06347},
  archivePrefix = {arXiv},
  primaryClass  = {cs.LG},
  url           = {https://arxiv.org/abs/1707.06347}
}

@article{higham2001algorithmic,
  author  = {Higham, Desmond J.},
  title   = {An Algorithmic Introduction to Numerical Simulation of Stochastic Differential Equations},
  journal = {SIAM Review},
  volume  = {43},
  number  = {3},
  pages   = {525--546},
  year    = {2001},
  doi     = {10.1137/S0036144500378302},
  url     = {https://doi.org/10.1137/S0036144500378302}
}

@misc{moallemi2022reinforcement,
  author = {Moallemi, Ciamac C. and Wang, Muye},
  title  = {A Reinforcement Learning Approach to Optimal Execution},
  year   = {2022},
  note   = {Working paper, current revision February 1, 2022}
}

@inproceedings{lin2020endtoend,
  author    = {Lin, Siyu and Beling, Peter A.},
  title     = {An End-to-End Optimal Trade Execution Framework based on Proximal Policy Optimization},
  booktitle = {Proceedings of the Twenty-Ninth International Joint Conference on Artificial Intelligence, {IJCAI-20}},
  publisher = {International Joint Conferences on Artificial Intelligence Organization},
  editor    = {Bessiere, Christian},
  pages     = {4548--4554},
  year      = {2020},
  note      = {Special Track on AI in FinTech},
  doi       = {10.24963/ijcai.2020/627},
  url       = {https://doi.org/10.24963/ijcai.2020/627}
}

@inproceedings{nevmyvaka2006reinforcement,
  author    = {Nevmyvaka, Yuriy and Feng, Yi and Kearns, Michael},
  title     = {Reinforcement Learning for Optimized Trade Execution},
  booktitle = {Proceedings of the 23rd International Conference on Machine Learning},
  pages     = {673--680},
  year      = {2006},
  publisher = {Association for Computing Machinery},
  doi       = {10.1145/1143844.1143929},
  url       = {https://doi.org/10.1145/1143844.1143929}
}

@misc{andrychowicz2020what,
  author        = {Andrychowicz, Marcin and Raichuk, Anton and Sta{\'n}czyk, Piotr and Orsini, Manu and Girgin, Sertan and Marinier, Rapha{\"e}l and Hussenot, L{\'e}onard and Geist, Matthieu and Pietquin, Olivier and Michalski, Marcin and Gelly, Sylvain and Bachem, Olivier},
  title         = {What Matters In On-Policy Reinforcement Learning? A Large-Scale Empirical Study},
  year          = {2020},
  eprint        = {2006.05990},
  archivePrefix = {arXiv},
  primaryClass  = {cs.LG},
  doi           = {10.48550/arXiv.2006.05990},
  url           = {https://arxiv.org/abs/2006.05990}
}

@article{wang2026reinforcement,
  author  = {Wang, Boyu and Gao, Xuefeng and Li, Lingfei},
  title   = {Reinforcement Learning for Continuous-Time Optimal Execution: Actor-Critic Algorithm and Error Analysis},
  journal = {Finance and Stochastics},
  volume  = {30},
  pages   = {597--655},
  year    = {2026},
  doi     = {10.1007/s00780-026-00589-5},
  url     = {https://doi.org/10.1007/s00780-026-00589-5}
}

@inproceedings{mania2019certainty,
  author    = {Mania, Horia and Tu, Stephen and Recht, Benjamin},
  title     = {Certainty Equivalence Is Efficient for Linear Quadratic Control},
  booktitle = {Advances in Neural Information Processing Systems},
  volume    = {32},
  year      = {2019},
  publisher = {Curran Associates, Inc.},
  url       = {https://proceedings.neurips.cc/paper/2019/hash/5dbc8390f17e019d300d5a162c3ce3bc-Abstract.html}
}

@misc{sowerby2022designing,
  author        = {Sowerby, Henry and Zhou, Zhiyuan and Littman, Michael L.},
  title         = {Designing Rewards for Fast Learning},
  year          = {2022},
  eprint        = {2205.15400},
  archivePrefix = {arXiv},
  primaryClass  = {cs.LG},
  doi           = {10.48550/arXiv.2205.15400},
  url           = {https://arxiv.org/abs/2205.15400}
}

@inproceedings{engstrom2020implementation,
  author    = {Engstrom, Logan and Ilyas, Andrew and Santurkar, Shibani and Tsipras, Dimitris and Janoos, Firdaus and Rudolph, Larry and Madry, Aleksander},
  title     = {Implementation Matters in Deep {RL}: A Case Study on {PPO} and {TRPO}},
  booktitle = {International Conference on Learning Representations},
  year      = {2020},
  url       = {https://openreview.net/forum?id=r1etN1rtPB}
}

@inproceedings{ng1999policy,
  author    = {Ng, Andrew Y. and Harada, Daishi and Russell, Stuart},
  title     = {Policy Invariance under Reward Transformations: Theory and Application to Reward Shaping},
  booktitle = {Proceedings of the 16th International Conference on Machine Learning},
  pages     = {278--287},
  year      = {1999},
  publisher = {Morgan Kaufmann},
  url       = {https://ai.stanford.edu/~ang/papers/icml99-shaping.pdf}
}

@inproceedings{hendricks2014reinforcement,
  author    = {Hendricks, Dieter and Wilcox, Diane},
  title     = {A Reinforcement Learning Extension to the {Almgren--Chriss} Framework for Optimal Trade Execution},
  booktitle = {2014 {IEEE} Conference on Computational Intelligence for Financial Engineering \& Economics ({CIFEr})},
  pages     = {457--464},
  year      = {2014},
  publisher = {IEEE},
  doi       = {10.1109/CIFEr.2014.6924109},
  url       = {https://doi.org/10.1109/CIFEr.2014.6924109}
}

@inproceedings{mukherjee2023bridging,
  author        = {Mukherjee, Amartya and Liu, Jun},
  title         = {Bridging Physics-Informed Neural Networks with Reinforcement Learning: {Hamilton--Jacobi--Bellman} Proximal Policy Optimization ({HJBPPO})},
  booktitle     = {{ICML} 2023 Workshop on New Frontiers in Learning, Control, and Dynamical Systems},
  year          = {2023},
  eprint        = {2302.00237},
  archivePrefix = {arXiv},
  primaryClass  = {cs.LG},
  url           = {https://arxiv.org/abs/2302.00237}
}

@inproceedings{johannink2019residual,
  author    = {Johannink, Tobias and Bahl, Shikhar and Nair, Ashvin and Luo, Jianlan and Kumar, Avinash and Loskyll, Matthias and Ojea, Juan Aparicio and Solowjow, Eugen and Levine, Sergey},
  title     = {Residual Reinforcement Learning for Robot Control},
  booktitle = {2019 International Conference on Robotics and Automation ({ICRA})},
  pages     = {6023--6029},
  year      = {2019},
  publisher = {IEEE},
  doi       = {10.1109/ICRA.2019.8794127},
  url       = {https://doi.org/10.1109/ICRA.2019.8794127}
}

@misc{macri2024reinforcement,
  author        = {Macr{\`i}, Andrea and Lillo, Fabrizio},
  title         = {Reinforcement Learning for Optimal Execution when Liquidity is Time-Varying},
  year          = {2024},
  eprint        = {2402.12049},
  archivePrefix = {arXiv},
  primaryClass  = {q-fin.TR},
  doi           = {10.48550/arXiv.2402.12049},
  url           = {https://arxiv.org/abs/2402.12049}
}

@inproceedings{fazel2018global,
  author    = {Fazel, Maryam and Ge, Rong and Kakade, Sham M. and Mesbahi, Mehran},
  title     = {Global Convergence of Policy Gradient Methods for the Linear Quadratic Regulator},
  booktitle = {Proceedings of the 35th International Conference on Machine Learning},
  series    = {Proceedings of Machine Learning Research},
  volume    = {80},
  pages     = {1467--1476},
  year      = {2018},
  publisher = {PMLR},
  url       = {https://proceedings.mlr.press/v80/fazel18a.html}
}

@inproceedings{henderson2018deep,
  author    = {Henderson, Peter and Islam, Riashat and Bachman, Philip and Pineau, Joelle and Precup, Doina and Meger, David},
  title     = {Deep Reinforcement Learning That Matters},
  booktitle = {Proceedings of the Thirty-Second AAAI Conference on Artificial Intelligence},
  volume    = {32},
  number    = {1},
  pages     = {3207--3214},
  year      = {2018},
  doi       = {10.1609/aaai.v32i1.11694},
  url       = {https://ojs.aaai.org/index.php/AAAI/article/view/11694}
}

\end{document}